\documentstyle[aps,prl,epsf,floats]{revtex}
\begin{document}
\twocolumn[\hsize\textwidth\columnwidth\hsize\csname@twocolumnfalse%
\endcsname
\title{Spin orthogonality catastrophe \\ in two-dimensional antiferromagnets and superconductors}
\author{Subir Sachdev$^1$, Matthias Troyer$^2$, and Matthias Vojta$^3$}
\address{$^1$Department of Physics, Yale University,
P.O. Box 208120, New Haven, CT 06520-8120, USA\\
$^2$Theoretische Physik, Eidgen\"ossische Technische Hochschule,
CH-8093 Z\"urich, Switzerland\\
$^3$Theoretische Physik III, Elektronische Korrelationen und
Magnetismus, Institut f\"ur Physik,\\ Universit\"at Augsburg,
D-86135 Augsburg, Germany}
\date{November 11, 2000}
\maketitle
\begin{abstract}
We compute the spectral function of a spin $S$ hole injected into
a two-dimensional antiferromagnet or superconductor in the
vicinity of a magnetic quantum critical point. We show that, near
van Hove singularities, the problem maps onto that of a static
vacancy carrying excess spin $S$. The hole creation operator is
characterized by a new {\em boundary\/} anomalous dimension and a
vanishing quasiparticle residue at the critical point. We discuss
possible relevance to photoemission spectra of cuprate
superconductors near the anti-nodal points.
\end{abstract}
\pacs{PACS numbers:}
]
% \newpage

A vigorous debate has been stimulated by photoemission
experiments\cite{zx} showing that the gapped {\em anti-nodal\/}
quasiparticle excitations of the high temperature superconductors
have a broad energy distribution curve (EDC). The many proposals
for this anomalous behavior include ({\em i\/}) electron
scattering by antiferromagnetic fluctuations \cite{afm}, ({\em
ii\/}) one-dimensional (1d) fluctuations at intermediate scales
leading to 1d electron fractionalization \cite{orgad}, ({\em
iii\/}) 2d electron fractionalization induced by proximity to
exotic 2d spin liquid states \cite{twod}, and ({\em iv\/})
coupling to superconducting phase and vortex fluctuations
\cite{phase}.

The studies of this paper fall into category ({\em i\/}); however,
our results also extend into a quasi 1d regime, and offer a
different perspective on category ({\em ii\/}). We will examine
the spectral function (or EDC) of the gapped anti-nodal
quasiparticles \cite{note} in the vicinity of a quantum critical
point between a $d$-wave superconductor and a state with
co-existing superconducting and antiferromagnetic order. Our
results will also apply to insulators in the vicinity of a
quantum transition between states with antiferromagnetic order
and a spin gap: the spectral functions are then those of holes
(and electrons) near van Hove singularities in the band structure
of the spin gap state. Our perspective on the superconducting
case differs from that of \cite{afm} in that we depart from
ground states in which pairing correlations have already induced
a gap to fermionic excitations in the anti-nodal regions, rather
than from states with a Fermi surface. Appealing to the proximity
of a magnetic quantum critical point between the former states
allows us to make controlled statements in a regime with strong
coupling between the fermionic quasiparticles (henceforth
referred to generically as holes) and the antiferromagnetic
fluctuations.

Our primary result is that the proper characterization of the
damping of the holes by low energy spin fluctuations is provided
by the framework of ``boundary'' critical phenomena \cite{cardy}.
The hole Green's function, $G_h$, is controlled by a new non-zero
boundary scaling dimension, $\eta_h$. Consequently, the hole
quasiparticle residue vanishes at the zero temperature ($T$)
quantum critical point due to an ``orthogonality catastrophe''
induced by a cloud of low energy spin excitations around the hole.
The motion of the hole charge does not play a dominant role: this
may be viewed as a form a ``spin-charge separation'', but is not
2d electron fractionalization in the sense of \cite{rs2,twod}.

As has been discussed elsewhere \cite{vbs}, spin fluctuations near
the magnetic quantum critical point in the superconductors or
insulators of interest are described by the following continuum
Hamiltonian
\begin{eqnarray}
{\cal H}_{\phi} &=& \int d^2 x \left[ \left( \pi_{\alpha}^2 +
c_1^2 (\partial_{x_1} \phi_{\alpha})^2 + c_2^2 (\partial_{x_2}
\phi_{\alpha})^2 \right)/2 \right. \nonumber \\
&+& \left. (r/2) \phi_{\alpha}^2 + (g/24) (\phi_{\alpha}^2)^2
\right] \label{h}
\end{eqnarray}
Here $\phi_{\alpha} (x,t)$ ($\alpha=1,2,3$) is the component of
the magnetization at wavevector ${\bf G}$ (usually ${\bf G} =
(\pi,\pi)$) at spatial co-ordinate $x$ and time $t$,
$\pi_{\alpha}$ is its canonically conjugate momentum, the only
non-zero equal-time commutation relation is $[\phi_{\alpha} (x,t),
\pi_{\beta} (x',t)] = i \delta_{\alpha\beta} \delta^2 (x-x')$,
$c_{1,2}$ are velocities (possibly unequal to account for a quasi
1d spatial anisotropy), $r$ is the parameter which tunes the
system across the quantum critical point ($\langle \phi_{\alpha}
\rangle \neq 0$ in the magnetically ordered phase for $r<r_c$, and
$\langle \phi_{\alpha} \rangle = 0$ otherwise), and $g$ is the
crucial quartic self interaction which is relevant below 3d and is
responsible for the non-trivial universal scaling properties of
the quantum critical point. There are no explicit damping terms
for $\phi_{\alpha}$ fluctuations in (\ref{h}) because the systems
of interest do not have a gapless particle-hole continuum near the
wavevector ${\bf G}$ \cite{dsc}.

We now inject a spin $S$ hole into the system described by ${\cal
H}_{\phi}$. We consider the hole spectrum near points of higher
symmetry in the Brillouin zone, with momentum ${\bf k} = {\bf
K}_1$, where its dispersion has a vanishing ${\bf k}$ derivative
{\em i.e.} near van Hove points like ${\bf K}_1 = (\pi, 0)$. We
denote the hole creation operator in the vicinity of ${\bf K}_1$
by $\psi_{1a}^{\dagger} (x,t)$, with $a=-S, \ldots, S$ a spin
index (for the case where the ground states are superconductors,
$\psi_{1a}^{\dagger}$ is the creation operator for the Bogoliubov
quasiparticles). Coupling to $\phi_{\alpha}$ fluctuations will
scatter the hole to momenta near ${\bf K}_2 = {\bf K}_1 + {\bf
G}$: we denote the hole creation operator near ${\bf K}_2$ by
$\psi_{2a}^{\dagger} (x,t)$, and it is assumed that ${\bf K}_2$
is also a van Hove point. The hole Hamiltonian is then
\begin{eqnarray}
{\cal H}_h &=& \int d^2 x \left[ \sum_{i=1,2} \left( \epsilon_i
\psi_{ia}^{\dagger} \psi_{i a} + \sum_{m=1,2} \alpha_{mi}
\psi_{ia}^{\dagger}
\partial_{x_m}^2 \psi_{ia} \right) \right. \nonumber \\
&+&  \gamma \phi_{\alpha} \left( \psi_{1 a}^{\dagger}
L_{ab}^{\alpha} \psi_{2 b} + \psi_{2 a}^{\dagger} L_{ab}^{\alpha}
\psi_{1 b} \right) \Biggr].
\end{eqnarray}
Here $L^{\alpha}$ are the familiar $(2S+1) \times (2S+1)$ angular
momentum matrices, the $\alpha_{mi}$ determine the band curvatures
near ${\bf K}_{1,2}$, and $\gamma$ is the coupling to the
$\phi_{\alpha}$ fluctuations. Higher order couplings between the
$\phi_{\alpha}$ and $\psi_{1,2}$ are also possible, but are
easily shown to be irrelevant under the renormalization group (RG)
discussed below. We assume that the hole energies, $\epsilon_i$
satisfy $\epsilon_1 = \epsilon_2 \equiv \epsilon_0>0$ (equality
holds with ${\bf K}_1=(\pi,0)$, ${\bf K}_2 = (0,\pi)$, and square
symmetry), for otherwise the hole scattering by $\phi_{\alpha}$
fluctuations is non-singular, and direct perturbative
computations of the hole spectrum are adequate. The singularity in
the value of $\epsilon_0$ at the magnetic transition at $r=r_c$
is weak and subdominant, and can be safely neglected.

A key observation follows from a simple, tree-level, RG analysis
of ${\cal H}_{\phi}+{\cal H}_h$. We know that ${\cal H}_{\phi}$ at
$r=r_c$ is invariant under the rescaling transformation $x
\rightarrow x e^{-\ell}$, $t \rightarrow t e^{-z \ell}$ with
$z=1$. Applying this to ${\cal H}_h$ we see immediately that the
$\alpha_{mi}$ flow as $d \alpha_{mi}/d\ell = - \alpha_{mi}$: so
the band curvatures are irrelevant for the low-energy theory, and
the hole may be viewed as dispersionless.

Before embarking on a complete RG analysis of ${\cal
H}_{\phi}+{\cal H}_h$, we make some qualitative observations on
the hole spectrum for $r<r_c$, $r>r_c$, and $r=r_c$.

For $r < r_c$, hole motion in the magnetically ordered state has
been studied earlier\cite{neel}. Because $\langle \phi_{\alpha}
\rangle \neq 0$, there is a non-zero mean matrix element between
the $\psi_{1a}$ and $\psi_{2a}$ states, and we have to
rediagonalize ${\cal H}_h$ to obtain the bare hole dispersion.
Simple considerations of energy and momentum conservation show
that there is an infinitely sharp quasiparticle pole in the
vicinity of the absolute band minimum: the slow quadratic
dispersion of the hole prevents decay by emission of linearly
dispersing spin waves. At frequencies, $\omega$, above this pole,
there is an incoherent gapless continuum, but its spectral weight
vanishes rapidly as $\omega$ approaches the quasiparticle pole:
after the rediagonalization of ${\cal H}_h$ required by a nonzero
$\langle \phi_{\alpha} \rangle$, it is easy to see that the
matrix element for emission of small momentum, Goldstone, spin
wave modes is suppressed by powers of the momentum. Similar
considerations also apply at other van Hove points which are not
global minima. However, depending upon non-universal details of
the band structure, in some cases it may be possible for the hole
to emit large wavevector, high energy $\phi_{\alpha}$ quanta, and
this would broaden the quasiparticle pole; for ${\cal H}_h$ such
processes occur if one of the $\alpha_{mi} < 0$, and require
momenta of order $c_{1,2}/\alpha_{mi}$ or larger (assuming ${\cal
H}_h$ still applies at such momenta). Given the irrelevance of the
$\alpha_{mi}$, the remainder of this paper will neglect this
non-universal decay. We will only consider low energy
$\phi_{\alpha}$ quanta, and assume that the high energy processes
are either not present, or contribute a small, background,
quasiparticle decay rate. If the latter were not true, there
would be no sharp quasiparticle-like peak or threshhold in the
hole spectrum, and the analysis of this paper would not be
necessary.

Closely related considerations apply in the spin-gap phase with
$r>r_c$ \cite{eder,sushkov}, but with some important differences:
({\em i\/}) the incoherent continuum is separated from the
quasiparticle pole by at least the spin gap energy, $\Delta \sim
(r - r_c)^{\nu}$, where $\nu$ is the correlation length exponent
of the phase transition in ${\cal H}_{\phi}$; ({\em ii\/}) the
$\phi_{\alpha}$ quanta are no longer Goldstone modes, and so the
matrix element for emission of a $\phi_{\alpha}$ quantum by the
hole does not vanish at zero momentum transfer.

The main purpose of this paper is to understand the nature of the
hole spectrum at the $T=0$ quantum critical point at $r=r_c$, and
its associated $T>0$ quantum-critical region. The $\phi_{\alpha}$
are now gapless critical excitations, but not Goldstone modes.
Consequently, there is no factor of a small momentum suppressing
their emission by the hole, and perturbative corrections in
$\gamma$ are infrared singular, as has also been noted by
Sushkov\cite{sushkov}. Our RG analysis will identify the
scale-invariant quantum field theory which permits a resummation
of the perturbative expansion, and shows that there is no
quasiparticle pole at $T=0$ and $r=r_c$; instead
\begin{equation}
G_h (\omega) = -{\cal A} \/ (\epsilon_0 - \omega)^{-1+\eta_h},
\label{orth}
\end{equation}
where ${\cal A}$ is a non-universal amplitude, and the universal
exponent $\eta_h$ is computed below.

For the critical theory, we can set $\alpha_{mi}=0$. Further,
examination of the perturbation theory in $\gamma$ shows that the
presence of two hole flavors, $i=1,2$, makes no material
difference to the critical singularities: the $\phi_{\alpha}
\rightarrow - \phi_{\alpha}$ symmetry of ${\cal H}_{\phi}$
ensures that the hole self-energy has terms only in even powers
of $\gamma$ for which the hole flavor returns to its original
value. Consequently we can drop the $i$ index, and refer to a
generic dispersionless hole $\psi_{a}$. After injection into the
antiferromagnet at $x=0$ (say), the $\psi_a$ charge will remain
localized at $x=0$, and its spin will couple to the
$\phi_{\alpha}$ fluctuations. So we are led to consider the
Bose-Kondo-like model \cite{ams,science,vbs}, ${\cal H}_{\phi} +
{\cal H}_S$ of a single quantum spin, $\hat{S}_{\alpha}$, coupled
to the bosonic $\phi_{\alpha}$ fluctuations where
\begin{equation}
{\cal H}_S = \gamma \hat{S}_{\alpha} \phi_{\alpha} (x=0),
\end{equation}
$[\hat{S}_{\alpha}, \hat{S}_{\beta}] = i
\epsilon_{\alpha\beta\gamma} \hat{S}_{\gamma}$, and
$\hat{S}_{\alpha} \hat{S}_{\alpha} = S(S+1)$. The charge density
of the injected hole can couple only to the spin-rotation
invariant $\phi_{\alpha}^2 (x=0)$, and such a term is irrelevant
under the RG \cite{vbs}.

The properties of ${\cal H}_{\phi} + {\cal H}_S$ have already
been studied in detail \cite{vbs} in the different physical
context of Zn/Li impurities in the cuprate superconductors; in
this earlier case the $\hat{S}_{\alpha}$ spin was permanently
confined near the impurity, while in the present situation there
is no impurity and $\hat{S}_{\alpha}$ is the spin of the injected
hole. Indeed, the relationship between ${\cal H}_{\phi} + {\cal
H}_S$ and ${\cal H}_{\phi} + {\cal H}_h$ is similar to that
between the familiar fermionic Kondo and X-ray edge problems.
However, the analogy is not perfect: it is conventional in the
X-ray edge problem to neglect the spin of the injected hole, and
merely couple its charge density to the fermionic bath. Here, the
spin exchange with the $\phi_{\alpha}$ quanta is paramount in
both cases.

The critical point of ${\cal H}_{\phi}$ defines a 2+1
dimensional, conformally invariant field theory, and ${\cal H}_S$
is a ``boundary'' perturbation along the line in spacetime at
$x=0$. This perturbation flows to a fixed point which is
invariant under conformal transformations which leave $x=0$
fixed. Correlations of $\hat{S}_{\alpha}$ are characterized by
its boundary anomalous dimension $\eta'/2$, which was computed to
two-loop order in an expansion in $\varepsilon=3-d$ in \cite{vbs}.

We are interested here in the $G_h = \langle \psi_a
\psi_a^{\dagger} \rangle$, and by (\ref{orth}) we identify
$\eta_h/2$ as the boundary anomalous dimension of $\psi_a$, which
is not simply related to $\eta'$. $G_h$ involves an overlap
between eigenstates of ${\cal H}_{\phi} + {\cal H}_S$ and states
in which the hole has been removed (the latter are outside the
Hilbert space of ${\cal H}_{\phi} + {\cal H}_S$). The needed
results do not follow from the previous analysis of ${\cal
H}_{\phi} + {\cal H}_S$ alone, and require extensions we describe
here.
%In particular, although $\hat{S}_{\alpha} = \psi_a^{\dagger}
%L_{ab}^{\alpha} \psi_b$, we have $\eta' \neq 2 \eta_h$: fusing the
%$\psi_a$, $\psi_a^{\dagger}$ at the same spacetime point induces a
%distinct anomalous dimension for the composite $\hat{S}_{\alpha}$
%operator.

\begin{figure}
\epsfxsize=1.5in \centerline{\epsffile{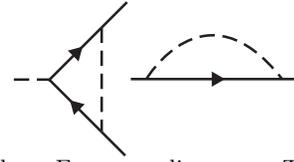}}
\caption{One-loop Feynman diagrams. The full line is $\psi_a$ and
the dashed line is $\phi_{\alpha}$.} \label{fig1}
\end{figure}

At one-loop order, we perform a standard momentum-shell RG of
${\cal H}_{\phi} + {\cal H}_h$ in $d$ spatial dimensions, in which
fields with momenta between $\Lambda$ and $\Lambda e^{-\ell}$, and
all frequencies, are integrated out. This is followed by the
rescalings $x \rightarrow x e^{-\ell}$, $t \rightarrow t e^{-
\ell}$, $\phi_{\alpha} \rightarrow \phi_{\alpha}
e^{(d-1+\eta)\ell/2} \phi_{\alpha}$, $\psi_a \rightarrow \psi_a
e^{(d+\eta_h)\ell/2}$. The RG flow of the bulk couplings $r$, $g$
is well known: $g$ approaches a finite fixed point value
$g^{\ast}$ at $r=r_c$, while $\eta=0$ at one loop order. For the
hole, evaluation of the one-loop graphs in Fig~\ref{fig1} shows
that $\eta_h = S(S+1) \widetilde{\gamma}^2$, where
$\widetilde{\gamma} = \gamma (c_1 c_2)^{-d/4}
\Lambda^{-\varepsilon/2} (\Gamma(d/2) (4 \pi)^{d/2})^{-1/2}$ obeys
the same flow equation as that obtained earlier for ${\cal
H}_{\phi}+{\cal H}_S$\cite{vbs,ams} (as expected):
\begin{equation}
d\widetilde{\gamma}/d\ell = \varepsilon \widetilde{\gamma}/2 -
\widetilde{\gamma}^3.
\end{equation}
We observe that $\widetilde{\gamma}$ also approaches a
fixed-point value, and at which we have the anomalous dimension
\begin{equation}
\eta_h = \varepsilon S(S+1)/2 + {\cal O}(\varepsilon^2).
\end{equation}
At the same order, a closely related computation shows that
$\eta'=\epsilon$ \cite{vbs}. The extension of these results on
the boundary exponents to two (and higher) loops requires the
field-theoretic renormalization group and the results are
presented elsewhere \cite{twoloop}: at next order there is
interference between bulk and boundary interactions, and the flow
equation for $\widetilde{\gamma}$ involves $g$. This interference
is a novel feature of the present problem and is absent in the
fermionic Kondo and X-ray edge problems, where the bulk degrees
of freedom can be represented by free fields. Direct numerical
evaluation of the two-loop corrections for the physical values
$\varepsilon=1$ and $S=1/2$ shows significant changes from the
one loop values: $\eta_h$ changes from $0.375$ to $0.087$, while
$\eta'$ changes from 1 to $0.232$.

\begin{figure}
\epsfxsize=3in \centerline{\epsffile{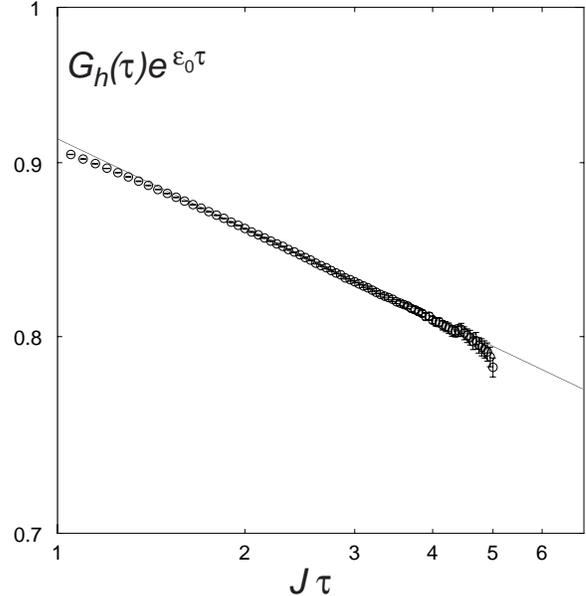}} \caption{Quantum
Monte Carlo results for a localized hole Green's function at the
quantum-critical point of a double layer Heisenberg
antiferromagnet. Both axes are logarithmic.} \label{fig3}
\end{figure}

We also computed $\eta_h$ from a continuous time world-line
quantum Monte Carlo \cite{loopalg} simulation of a double layer
Heisenberg antiferromagnet at its quantum critical point
\cite{doublelayer}. We measured $G_h (\tau)$ ($\tau$ is imaginary
time) for a single hole on a fixed site $i$, by relating it to
correlators of spin world lines at site $i$ which do not flip in
time $\tau$. We obtained the estimate $\eta_h = 0.087 \pm 0.040$
from simulations on a $64\times 64$ system by fitting to
$G_h(\tau)\sim \tau^{-\eta_{h}}\exp(-\epsilon_0\tau)$, in the
range $2 \leq J \tau \leq 6$ ($J$ is the intralayer exchange, and
$\epsilon_0/J =2.102 \pm 0.016$ was determined separately from the
ground state energies of the antiferromagnet with and without a
hole); the numerical data is in Fig~\ref{fig3}.

\begin{figure}
\epsfxsize=3in \centerline{\epsffile{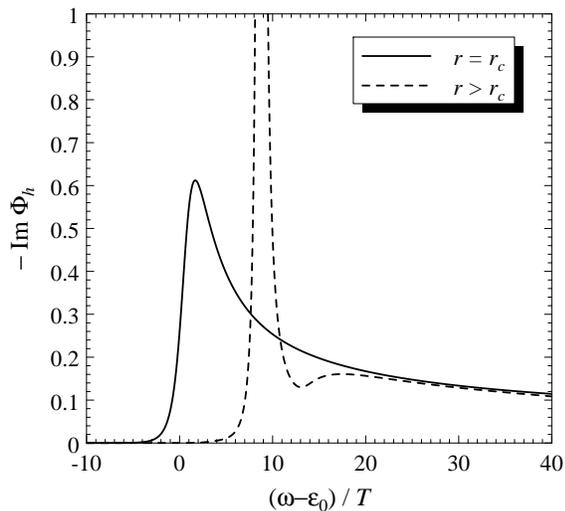}} \caption{Large $N$
\protect\cite{vbs} results for the imaginary part of the scaling
function $\Phi_h$ in (\protect\ref{gscale}). At large $N$
$\eta_h=1/2$. We used an interpolation form for the
$\phi_{\alpha}$ propagator $1/(k^2 - \omega^2 + m^2 - 2 i \Gamma
\omega)$, where $m/T$ and $\Gamma/T$ are universal functions of
$\Delta/T$ \protect\cite{book}. Both $m$ and the damping,
$\Gamma$, are non-zero even at $r=r_c$ because the thermally
excited $\phi_{\alpha}$ quanta scatter strongly off each other by
the non-zero fixed point value of the bulk interaction, $g$, in
${\cal H}_{\phi}$. For $r=r_c$ we used $m=\Gamma=T$, and for
$r>r_c$ we used $\Gamma=m/5=T$.} \label{fig2}
\end{figure}

In the $r \geq r_c$, $T \geq 0$ vicinity of the critical point,
(\ref{orth}) generalizes to
\begin{equation}
G_h(\omega) = \frac{{\cal A}}{T^{1-\eta_h}} \Phi_h \left(
\frac{\omega-\epsilon_0}{T}, \frac{\Delta}{T} \right)
\label{gscale}
\end{equation}
where ${\cal A}$ and $\epsilon_0$ are the same constants appearing
in (\ref{orth}) (they have absorbed a non-universal
renormalization from the coupling to $\phi_{\alpha}$ modes), while
$\Phi_h$ is a completely universal function. We obtained numerical
results for $\Phi_h$ using the large $N$ method discussed in
\cite{vbs} and the results are shown in Fig~\ref{fig2}. The ${\bf
k}$ dependence of $G_h$ arises only from the irrelevant
$\alpha_{mi}$ couplings, and their main effect is to replace
$\epsilon_0$ by the actual hole dispersion near the van-Hove
point.

We conclude by discussing possible physical application of our
results to photoemission measurements on the cuprate
superconductors. We have already argued elsewhere
\cite{vbs,scienceqpt} that (\ref{h}) should be a reasonable
description of the antiferromagnetic fluctuations in the low and
moderate doping regime, both above and below the superconducting
$T_c$. There is evidence from NMR \cite{imai} and neutron
scattering experiments \cite{gabe} that the quantum critical
region of the $r=r_c$ critical point describes the
antiferromagnetic fluctuations above $T_c$; so for the same
systems, the $r=r_c$ spectrum in Fig~\ref{fig2} should apply to
photoemission at the anti-nodal points. The proximity of such a
magnetic quantum critical point may be associated with the onset
of quasi 1d correlations, but this incidental to our theory---the
anisotropy is merely reflected by changes to the couplings in
${\cal H}_{\phi}$. Below $T_c$, the measured anti-nodal spectrum
\cite{ding} is similar to the $r>r_c$ spectrum in Fig~\ref{fig2}:
this is accounted for in our approach by the reasonable assumption
that the onset of superconductivity induces the spin-gap-like
correlations and so increases the value of the effective $r$
controlling the magnetic fluctuations. The high frequency tail of
the EDC both above and below $T_c$ should decay as
$1/\omega^{1-\eta_h}$. Note that this tail is present for any
non-zero value of $\eta_h$, with a non-universal amplitude
determined by ${\cal A}$. However, our estimate of the value of
$\eta_h$ here is quite small, and it appears that the interaction
effects discussed here cannot explain the prominent tail seen in
current experiments on their own. It is likely that the strong
disorder in the local gap values (as seen in recent STM
measurements \cite{seamus}) also plays an important role.

We thank A.~Castro Neto, A.~Chubukov, V.~Kotov, and J.~Schmalian
for useful discussions. This research was supported by NSF Grant
DMR 96--23181 (USA), the Swiss NSF, and SFB 484 (Germany).

\begin{references}
\bibitem{dsc} In $d$-wave superconductors, there is a gapless
particle hole continuum arising from the nodal Fermi points, but
this is at the wavevector, ${\bf K}$, separating two nodal
points, and it is assumed that ${\bf K} \neq {\bf G}$. The case
${\bf K}={\bf G}$ can also be analyzed by the methods of this
paper: the results are very similar \protect\cite{vbs}, and the
value of $\eta_h$ is modified only at the two loop level.
\end{references}
\end{document}